\documentclass[reprint, amsmath,amssymb, aps,pra, groupedaddress]{revtex4-1}
\usepackage{graphicx}% Include figure files
\usepackage{dcolumn}% Align table columns on decimal point
\usepackage{bm}% bold math

\newcommand{\micron}{\,\ensuremath{\mu{\rm m}}}
\newcommand{\microsecond}{\,\ensuremath{\mu{\rm s}}}
\newcommand{\ohmm}{\,\ensuremath{\Omega\cdot{\rm m}} }

\newcommand{\bra}[1]{\langle #1|}
\newcommand{\ket}[1]{|#1\rangle}
\newcommand{\w}{3.3in}

\begin{document}

\title{Half Cycle Pulse Train Induced State Redistribution of Rydberg Atoms}

\author{P. K. Mandal}
\author{A. Speck}
 \email{speck@rowland.harvard.edu}
\affiliation{The Rowland Institute at Harvard, Cambridge, MA 02142, USA}

\date{\today}

\begin{abstract}
Population transfer between low lying Rydberg states independent of the initial state is realized using a train of half-cycle pulses with pulse durations much less than the classical orbit period.  We demonstrate experimentally the transfer of population from initial states around $n=50$ with 10\% of the population de-excited down to $n < 40$ as well as up to the continuum.  This is the first demonstration of a state-independent de-excitation technique applicable to the currently produced state distribution of antihydrogen. The measured population transfer matches well to a model of the process for 1-D atoms.
\end{abstract}

% PACS, the Physics and Astronomy
                             % Classification Scheme.
\maketitle

\newcommand{\ApparatusFigure}{
\begin{figure}[b]
\centering
\includegraphics*[width=\w]{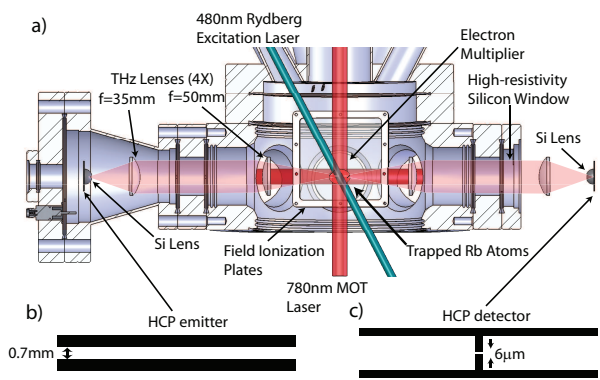}
\caption{Schematic of the apparatus.  Atoms trapped in the center of the magneto-optic trap are excited to a Rydberg state, interact with a train of half-cycle pulses generated by the emitter antenna on the left side, and then are detected by field ionization from the field plates surrounding the central trapping region.   The emitter and photoconductive detector geometries are shown in (b) and (c) respectively.} \label{fig:apparatus}
\end{figure}
}

\newcommand{\HCPFigure}{
\begin{figure}
\centering
\includegraphics*[width=\w]{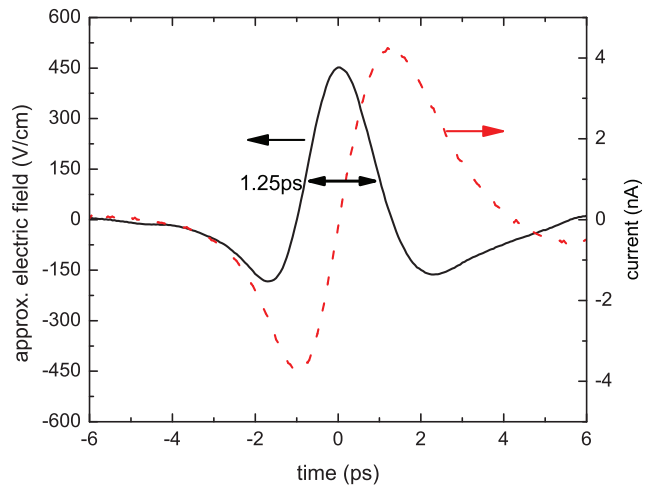}
\caption{ Half-cycle pulse waveform for a biasing electric field of 2 kV/cm.  The dashed line shows the measured current from the photoconductive detector while the solid line is the derivative of the current which is proportional to the radiated electric field.} \label{fig:hcp}
\end{figure}
}

\newcommand{\ContourFigure}{
\begin{figure}
\centering
\includegraphics*[width=\w]{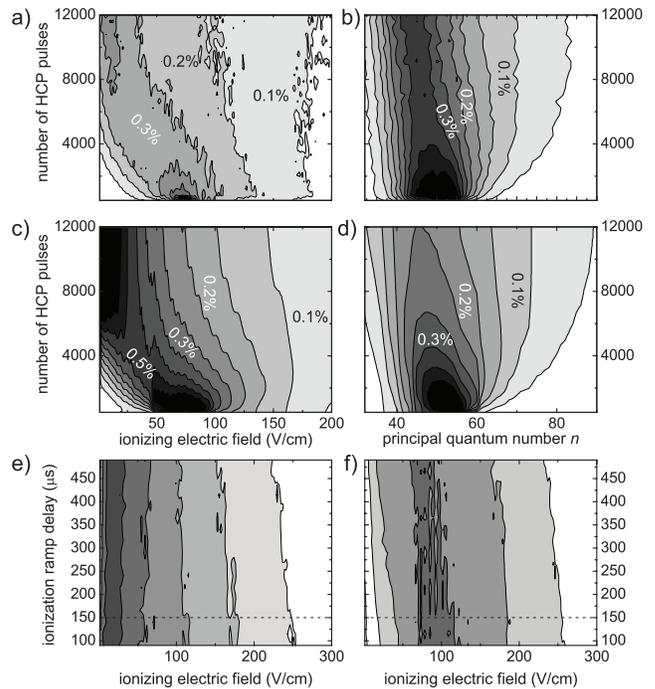}
\caption{Left plots: Normalized fraction of the atomic population that ionizes at a specific electric field as a function of the number of applied half-cycle pulses. (a) shows experimental data and (c) shows the calculated values for the 1-D model with a momentum transferred per pulse of $q=7.9\times 10^{-5}$ in atomic units.  Right plots: Atomic state distribution based on conversion from measured ionizing field using the expected fields for red-shifted Stark states.  (b) shows the experimental data while (d) is calculated using the 1-D model. Bottom plots: Evolution of the state distribution in time after 6500 (e) or 2400 (f) HCP's are applied.  The evolution is measured by delaying the ionizing ramp relative to the initial Rydberg excitation pulse. The dashed lines shows the ionization ramp delay of 150\microsecond\,used for all other plots.   } \label{fig:contour}
\end{figure}
}

\newcommand{\MedianFigure}[1][\tw]{
\begin{figure}
\centering
\includegraphics*[width=\w]{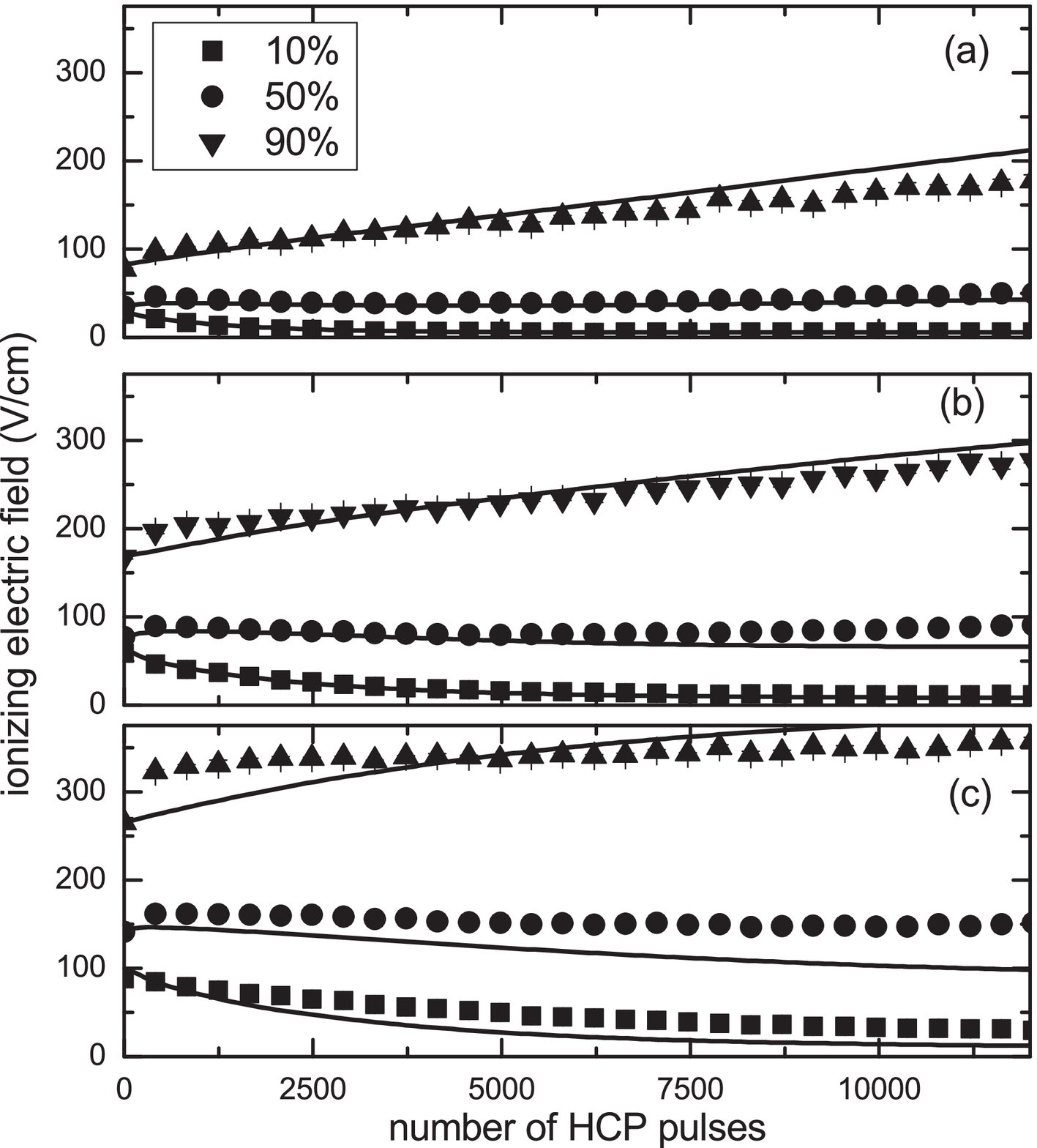}
\caption{Electric field for which the given percentage of the atomic ensemble has ionized as a function of the number of half-cycle pulses applied.  The initial state of the atoms varies from (a) $n=62$, (b) $n=52$, to (c) $n=46$.  Solid lines show the expected value from the 1-D model with momentum transfer per pulse of $q=7.9\times 10^{-5}$ in atomic units.}\label{fig:median}
\end{figure}
}

\newcommand{\QHCPFigure}{
\begin{figure}
\centering
\includegraphics*[width=\w]{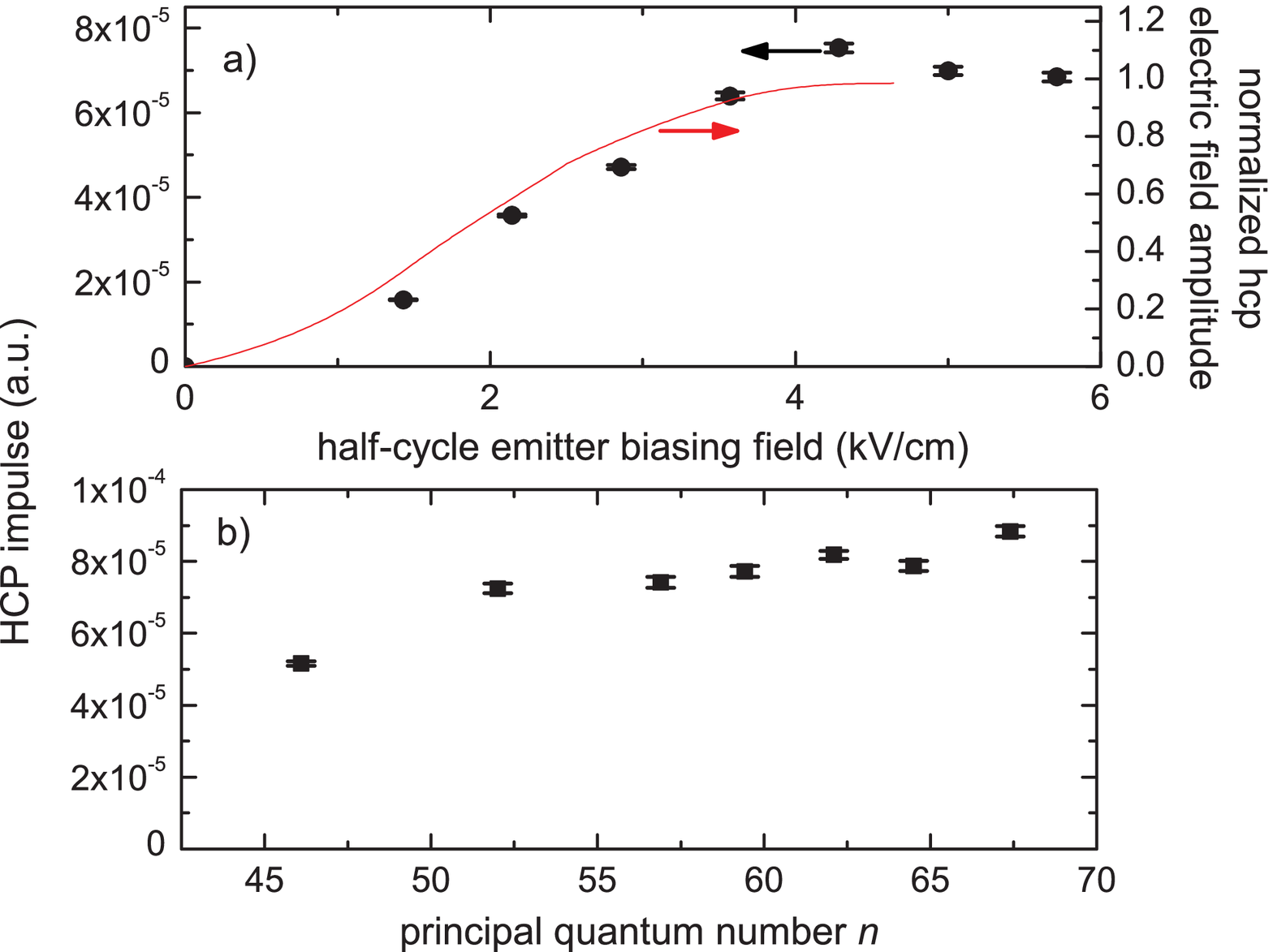}
\caption{Best fit for the impulse in atomic units transferred by a single half-cycle pulse to the 1-D model described in the text.  (a) the initial state of the atoms was held fixed at $n=62$ while the amplitude of the half-cycle pulse was scanned.  The solid line shows the measured amplitude of the half-cycle pulse.  (b) the half cycle amplitude was held constant at a biasing field of 4.3 kV/cm while the initial atomic state was varied.  } \label{fig:qhcp}
\end{figure}
}

\newcommand{\REDNCalFigure}{
\begin{figure}
\centering
\includegraphics*[width=\w]{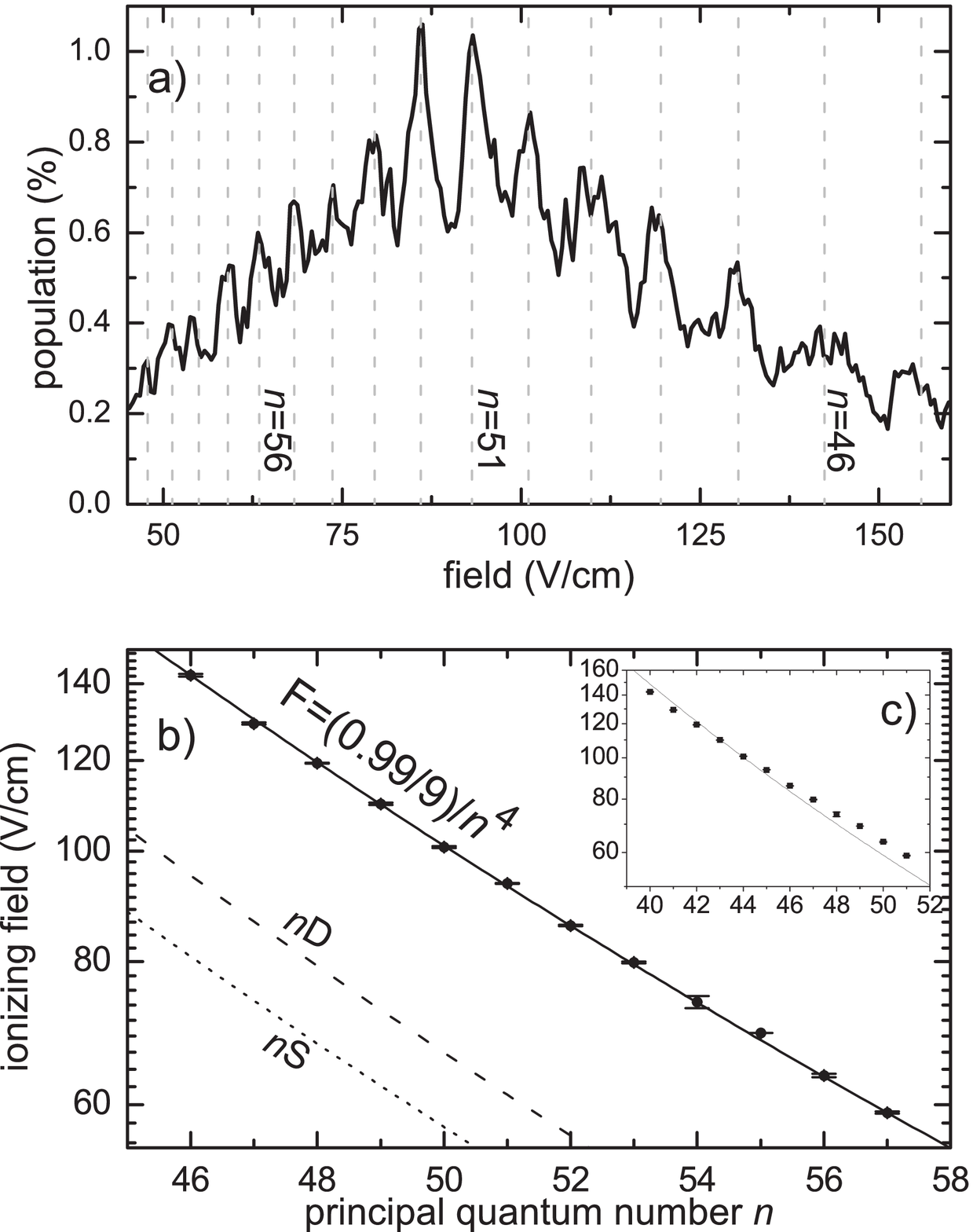}
\caption{(a) Field ionization trace showing peaks from populations in many different principal quantum numbers.  Dashed lines are the calculated ionizing field for red Stark states.  (b) Ionization peak locations versus principal quantum number.  Note the close agreement to the red Stark state calculated value of $F=(1/9)/n^4$. The dashed and dotted lines shows the expected ionization fields for \textit{n}S and \textit{n}D states, respectively, following adiabatic pathways towards ionization. (c) Similar fit assuming adiabatic ionization of \textit{n}D states. } \label{fig:redncal}
\end{figure}
}

\newcommand{\TransProbFigure}{
\begin{figure}
\centering
\includegraphics*[width=\w]{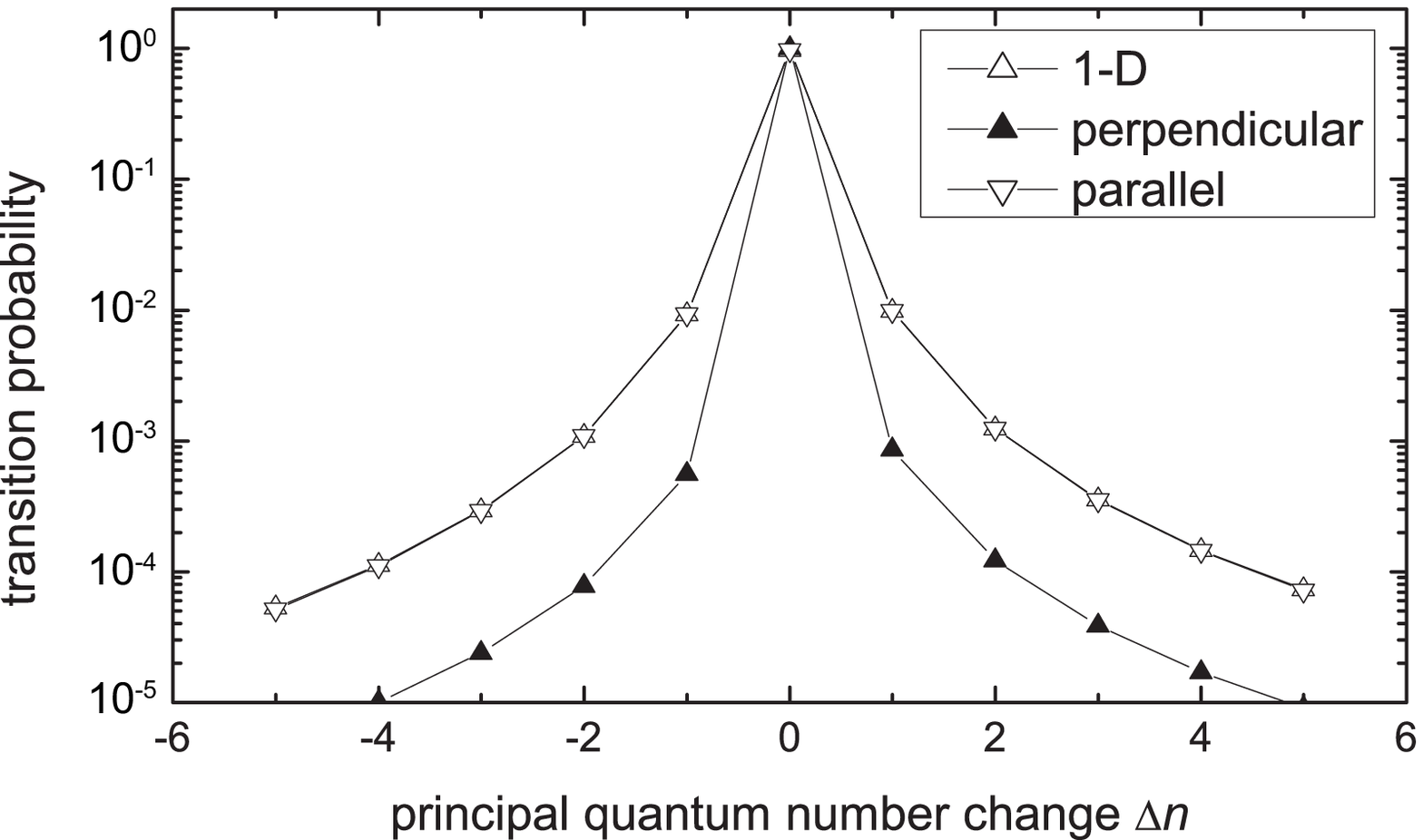}
\caption{Transition probability from $n=62$ to $n=62+\Delta n$ for a momentum transfer of $q=7.9\times 10^{-5}$.  The probabilities from the 1-D model (Eq.  \ref{eq:t}) and from 3-D red-shifted Stark states with $| m | \leq 5 $ kicked by a pulse with electric field polarization along the quantization axis are almost the same.  The probabilities from the same Stark states kicked instead by a pulse with electric field polarization perpendicular to the quantization axis are more than an order of magnitude lower.} \label{fig:tprob}
\end{figure}
}

\section{Introduction}

The control of atoms in Rydberg states is important for many different applications including the study of anti-hydrogen, the study of wavepackets in Rydberg atoms, and the use of excitation blockades for quantum computing\cite{Singer2004,Tong2004,Jones1998}.  Precise tests of CPT using antihydrogen requires trapped atoms in the ground state, but currently atoms are produced in a broad range of excited states.  An effective means of de-excitation is thus required to ensure atoms reach the ground state and are trapped before annihilating on the trap walls.  Stimulated de-excitation using narrow-band lasers is clearly infeasible due to the many thousands of states populated during three-body recombination.  Hence an alternative state-independent technique is required.  Efficient de-excitation of anti-hydrogen atoms compatible with the techniques used currently to produce anti-hydrogen requires a scheme that meets several requirements.  First it must be capable of simultaneously de-exciting a range of states whose radius varies from 0.5\micron\, to less than 100 nm; this range is equivalent to states in no magnetic field with principal quantum numbers, $n$, between $n<35$ and $n=90$\cite{Gabrielse2002}.  Second the current techniques to produce antihydrogen atoms result in a quasi-static production rate over at least several seconds; a de-excitation scheme must thus not require a specific timing sequence.  Lastly, atoms with thermal velocities comparable to the 1.2K production environment reach the walls of the trap enclosure in under 250\microsecond\, which sets a limit for the required de-excitation rate.

Several groups have proposed schemes using trains of ultra-short unipolar electromagnetic pulses (half-cycle pulses or HCP) to accomplish this de-excitation.   When the length of the pulse is much shorter than the orbit period of the electron, $\tau_{orbit} \propto n^{3}$, these pulses effectively kick the electron transferring momentum regardless of the initial state\cite{Jones1993}.  Depending on the phase of the kick relative to the electron's orbit this momentum transfer can either drive the atomic state up towards the continuum or down towards the ground state. Hu and Collins solved the applicable time-dependent Schr\"{o}dinger equation which predicts that a fixed repetition rate train of pulses should result in an oscillating population distribution in number of half-cycle pulses with excursions from the initial state of over $\Delta n = \pm 10$\cite{Hu2004}.  Kopyciuk and Parzynski noted that, in a model system based on 1-D atoms, by changing the repetition rate of the half-cycle pulses during an experimental sequence atoms can be driven either up or down in principal quantum by decreasing or increasing the repetition rate respectively. Their model however does not predict the oscillatory phenomenon seen by Hu and Collins\cite{Kopyciuk2009,Kopyciuk2007}.

Experimentally, half-cycle pulse trains have been demonstrated to coherently transfer Rydberg atoms between \textit{n}-states of 350 and 700\cite{Mestayer2007}.  As the orbit period for these atoms is on the order of tens of nanoseconds, electrical impulse generators are used to generate pulses with temporal widths less than the orbit period.  However this experimental technique cannot be extended to atoms with $n < 100$ due to the much shorter pulse lengths required when the orbit periods are less than 100 ps.  Single pulses to atoms in states comparable to those produced for antihydrogen study have been shown to cause population redistribution to nearby \textit{n} and \textit{l} states but ionization occurs at high pulse amplitudes instead of inducing larger changes in quantum state\cite{Wesdorp2001,Tielking1995}.   These systems utilize amplified femtosecond lasers with repetition rates limited to less than around 10 kHz to produce half-cycle pulses; they thus cannot be extended to produce pulse trains to drive large changes within the lifetime of a Rydberg atom.  Utilizing adiabatic rapid passage through a series of microwave resonances has also been demonstrated as a technique to drive population transfers but requires specific knowledge of the dressed atomic states which is difficult in strong inhomogeneous magnetic fields such as those used to trap antihydrogen\cite{Lambert2002,Maeda2006}.  

In this paper we report on the first experimental demonstration of a technique that meets all of the requirements for efficient de-excitation.  An 82 MHz repetition rate photoconductive switch produces a train of primarily unipolar pulses with pulse widths on the order of 1 ps and amplitudes greater than 500 V/cm allowing for the manipulation of atoms with $n>20$ through impulsive kicks from these pulses.  By applying thousands of pulses within the radiative lifetime of an atom, $\Gamma \sim 100 \microsecond$ for $n \sim 40$, over 10\% of the initial population in $n = 52$ can be driven to $n = 40$.  As radiative lifetimes scale strongly with $n$, $\Gamma_{\mathrm{rad}} \propto n^5/\log n$, this reduction in \textit{n} results in a factor of 4 reduction in lifetime\cite{Chang}.  The population transfer follows the predictions of the 1-D model system used in the theoretical work but instead of a coherent transfer from each pulse the population decoheres after each pulse.

\section{Apparatus}

Fig. \ref{fig:apparatus}(a) shows the primary apparatus used in this experiment.  Approximately $3 \times 10^{8}$ $^{85}\mathrm{Rb}$ atoms are trapped in a 4 mm diameter cloud in the center of a dispenser loaded magneto-optical trap.  The combination of the primary cooling laser at 780 nm and a tunable 480 nm optical parametric oscillator producing 2 mJ, 10 ns long pulses with linewidth of 4 $\mathrm{cm}^{-1}$ excite 1 in $10^{4}$ atoms in two steps from the 5S ground state to Rydberg \textit{n}S or \textit{n}D states.  Field plates of 80\% transparent stainless steel mesh separated by 2.2 cm can be biased producing an electric field that field ionizes the Rydberg atoms.  An electron multiplier detects these ionized electrons.  By applying a linear ramped field in time the entire field ionization distribution can be mapped out in a single shot.  This field typically ramps from 0 to 700 V/cm in 10\microsecond.

\ApparatusFigure

Half cycle pulses are generated within the vacuum system using a photoconductive switch fabricated on a semi-insulating GaAs wafer with micro-fabrication techniques\cite{Zhao2002}. SI-GaAs is a slow photoconductive switch with a carrier lifetime of more than 100 ps, a characteristic which is necessary for producing half-cycle pulses (HCP) as we will discuss later. The semiconductor wafers used here are cut along the $\mathit{\langle 100 \rangle}$ direction and have a resistivity of $10^{7} \ohmm$. Aluminum strip lines with a gap of 700\micron\, for biasing the switch are directly deposited on GaAs using electron-beam evaporation to a thickness of between 0.5\micron\, and 0.8\micron.  Fig. \ref{fig:apparatus}(b) shows the basic geometry.  A similar wafer is used as a detector but instead of a stripline antenna a dipole antenna with a gap of 6 \micron\, is fabricated on the detector as shown in Fig. \ref{fig:apparatus}(c).  A femtosecond Ti:sapphire oscillator with a maximum pulse width of 100\,fs illuminates the emitter and detectors. This mode-locked laser operates around a center wavelength of 810 nm with a repetition rate of 82\,MHz. An average laser power as high as 450\,mW focused to a spot size of 50\micron\, was used to pump the emitter.

To compensate for the strongly diverging nature of the emitted HCP radiation, a combination of a 10 mm diameter hemispherical silicon lens  located on the back of the emitter wafer along with a pair of plano-convex lenses are used to refocus the pulses at the center of the trapped atom cloud.  The two lenses have focal lengths of 35 mm and then 50 mm and are respectively composed of TPX (polymethylpentene) and high-density polyethylene which are both transparent in the terahertz region. They are placed such that all frequencies are refocused at the center of the trapped atoms.  An identical sequence of lenses directs and refocuses the outgoing pulses on the photoconductive detector located outside the vacuum system.  A high-resistivity Si window allows the HCP pulses to exit the UHV system.

\HCPFigure

Fig. \ref{fig:hcp} shows typical half-cycle pulses with a temporal FWHM of approximately 1 ps and a peak field of over 500 V/cm produced by our setup.  The fast excitation of carriers in the GaAs wafer due to the pump pulse coupled with the strong bias electric field results in a sudden current surge.  Since in the dipole approximation the far field radiation is proportional to the time derivative of the current, this current surge produces a strong positive pulse while the long carrier lifetime causes a slow decay in the current leading to a long negative tail.  After propagation in free space, the strong divergence of the low frequency components of the pulse as well as the Gouy phase induced by focusing converts this pulse into the symmetric pulse shown in Fig. \ref{fig:hcp}.  

The pulses are detected using the inverse technique in which the half-cycle pulse electric field is used to induce a current within the detector.  The measured current density is then given by $j = e\mu \int_{-\infty}^{\infty} E_{HCP}(t) n(t) dt$ where $e$ is the charge of the carriers, $\mu$ is the carrier mobility, and $n(t)$ is the carrier density.  For a semiconductor like SI-GaAs with a long carrier lifetime, $n(t)$ can be approximated by a step function at time $\tau$, corresponding to the arrival of the probe pulse, so this equation reduces to $j(\tau) \propto \int_{\tau}^{\infty} E_{HCP}(t-\tau) dt$ or equivalently the electric field of the half-cycle pulse is proportional to the derivative of the measured current.  Using the simplifying assumption that the carrier mobility at terahertz frequencies is the same as that at DC frequencies and accounting for both the exponential decay in carrier density caused by the approximately 300 ps carrier lifetime of SI-GaAs and the reflection losses at the interface between the detector and free space, the current induced by a known DC biasing field is used to calibrate the detector.  This calibration has been checked by comparing this detected signal with an electro-optic detector with a known absolute calibration.

\ContourFigure

Each experimental run is synchronized relative to a laser pulse from the OPO which provides the excitation to a given Rydberg state.  3 ms before the pulse the trapping quadrupole magnetic field is turned off allowing all resulting eddy currents to dissipate prior to the excitation and ensuring minimal interference from the quadrupole field.  5\microsecond\, prior to the pulse the photoconductive switch's bias field is switched on allowing it to stabilize prior to any Rydberg atom excitation.  This eliminates any effects from lower amplitude pulses being emitted as the field is being raised to the full value.  Approximately 30,000 atoms are excited to a Rydberg state at time $t=0$.  After a variable time between 0 and 150\microsecond, the half-cycle pulse biasing field is turned off with a fall time of 30 ns.  The final state distribution is measured by applying a 10\microsecond\, long ionizing field ramp at a fixed time of 150\microsecond\, after the excitation pulse.  By fixing the ionization field ramp in time relative to the excitation pulse, we ensure that no effects due to radiative decay or transitions induced by blackbody radiation are convolved into the measured state distribution because the only change between experimental runs is the length of time the half-cycle pulses are applied.  Fig. \ref{fig:contour} (e) and (f) show the measured normalized distribution for a fixed number of half-cycle pulses as the delay between the Rydberg excitation laser pulse and the ionization ramp is varied.  At long delay times there is a small decrease in the number of atoms ionized at high fields as expected due to these states having reduced lifetimes but at a delay of 150\microsecond, used for all of our other data, there is a negligible effect on the overall state distribution.

\REDNCalFigure

\section{Results and Discussion}

Fig. \ref{fig:contour}(a) shows the normalized state distribution measured as a function of the number of half-cycle pulses.  Initially all atoms ionize near 80 V/cm as set by the excitation wavelength and resulting state.  As impulsive kicks are applied to this initial distribution the population spreads out in a diffusive manner. During this process some atoms are lost due to field ionization by the small static 1V/cm clearing field or from radiative decay to states that are not ionized by the maximum field of 700 V/cm applied during the detection ramp.  To map between ionizing electric field and the approximate principal quantum number distribution shown in Fig. \ref{fig:contour}(c) we utilize the electric field at with the electron is no longer classically bound.  In atomic units this field is $F=\frac{A}{n^4}$. Here $A=1/16$ for adiabatic ionization or $A=1/9$ for red-shifted Stark states that diabatically ionize which we employ for data analysis from now on\cite{Gallagher1994}.  Justification for the choice of red-shifted Stark states is shown in Fig. \ref{fig:redncal}.  Part (a) shows a field ionization trace after 1600 half-cycle pulses have been applied to the population.  More than 10 peaks corresponding to discrete atomic states are clearly visible.  Part (b) plots these peak positions versus the best fit principal quantum number and corresponds closely to a distribution from red-shifted Stark states.  The inset shows a comparison to the expected distribution if we assume \textit{n}D states have been produced which does not agree as well.  The predominant excitation of red-shifted Stark states is due to the small electric field present during excitation\cite{Stokely2003}.  Due to the large quantum defects of $l \leq 2$ states in Rb, these optically accessible states are displaced at zero field from the Stark manifold of states with $l > 2$.  While the first Stark manifold that a \textit{n}S state encounters is the $n-3$ manifold with which there is only a small coupling due to the large difference in \textit{n}, a \textit{n}D state encounters the downhill, red-shifted state of the $n-1$ Stark manifold where there is a much larger coupling.  At this point much of the \textit{D} character is spread to these red-shifted states and it becomes possible to optically excite them when the electric field is high enough to have reached the first crossing, $F \sim 1/3n^5$.  

Assuming the length of the half-cycle pulse is short relative to the electron orbit time, the transition amplitude from initial state, $i$, to final state, $f$, due to an interaction with a HCP can be reduced to the inelastic form factor of the atom, $T_{f,i}=\bra{f}e^{-i \vec{q} \cdot \vec{r}}\ket{i}$ where $\vec{q}$ is the momentum transferred to the atom.  With an initial density matrix, $\rho(0)$, given by a statistical mixture of the measured initial state distribution, the state distribution resulting after $j$ pulses is given by
\begin{equation}
\rho(j) = \left ( e^{-iHt_{r}}T\right)^{j} \rho(0) \left ( T^{\dagger} e^{iHt_{r}}\right)^{j}
\end{equation}
where $t_{r}$ is the period of the pulse train and $H=-1/2n^2$ is the unperturbed Hamiltonian of the system.  

The applicable form factor has been analytically calculated for hydrogenic Stark states with parabolic quantum numbers $n_{1}$, $n_{2}$, $m$, and $n = n_{1} + n_{2} + |m| + 1$ where the atomic quantization axis and the HCP electric field polarization are parallel\cite{Bersons1997}.  In this configuration the only applicable selection rule is $\Delta m = 0$.  When the quantization axis and the HCP electric field polarization are perpendicular, there is currently no simple analytical formula.  In the case of weak momentum transfer, $T_{f,i} \sim \delta_{i,f} - iq_{x}\bra{f}x\ket{i}$ and the applicable matrix elements are known\cite{Dewangan2005}.  In contrast to the parallel case, the selection rule for the magnetic quantum number is now $\Delta m = \pm 1$.  As red-shifted Stark states are predominantly one-dimensional atoms with the expectation value of the axial electron position much greater than the radial position, the interaction can be further simplified to that of 1-D atoms interacting with an impulsive kick of given momentum transfer, $q$.  The transition amplitude for a single impulsive kick in the 1-D approximation is given by\cite{Bersons2004}:
\begin{multline}
T_{n',n}=-\frac{z\left(\lambda-2/n\right)^{n}\left(\lambda-2/n'\right)^{n'}}{\sqrt{nn'}\lambda^{n+n'}}  \Biggl\{\biggl[\frac{n-1}{\lambda-2/n} \label{eq:t} \\
+\frac{n'-1}{\lambda-2/n'}-\frac{n+n'}{\lambda}\biggr]F(-n+1,-n'+1,2,z) \\
-\frac{(n-1)(n'-1)}{2}\biggl[\frac{1}{\lambda-2/n}+\frac{1}{\lambda-2/n'}\biggr]  \\
\times zF(-n+2,-n'+2,3,z)\Biggr\}
\end{multline}
where $n$ and $n'$ are the initial and final principal quantum numbers, $F(a,b,c,y)$ is the hypergeometric function, $\lambda = n^{-1} + (n')^{-1}-iq$, and $z = -4n'n\left[(n-n')^{2}+(qn'n)^{2}\right]^{-1}$.  

Fig. \ref{fig:tprob} compares the probability, $|T_{f,i}|^{2}$, of a transition for these three different cases from a red-shifted Stark state where $n_{1}=0$ and $|m|\leq 5$ to a final state with a given change in principal quantum number.  Two points from this figure are important.  First the overall shape of the curve is similar for all three cases but the probability of a transition in the perpendicular kick case is approximately an order of magnitude lower.  Since $\Delta E = (2p_{0}\cdot q + q^{2})/2$ where $p_{0}$ is the initial momentum and with perpendicular polarization $p_{0}\cdot q = 0$, one would expect the reduction in energy transfer and thus transition amplitude when kicked in this configuration.   Second, in the two 3-D cases the transition amplitude is clustered around similar states.  Hence the likelihood of driving a transition away from red-shifted Stark states is many orders of magnitude lower.  Given these results, the 1-D model is expected to apply quite closely apart from the correction in overall amplitude necessary for the perpendicular kick case as atoms that start in a primarily 1-D state will stay in similar states.

\TransProbFigure

Unless a fast decoherence term is added, this model shows interference effects and emphasizes eigenstates which have stationary phase at the frequency of the pulse train.  This is not seen in the actual experimental data suggesting that decoherence occurs on a time scale similar to the time between half-cycle pulses.  This is a shorter time scale than dephasing due to blackbody transitions or spontaneous emission.  Also no dependence on density is observed precluding a collisional process being responsible for the decoherence.  The most probable reason for the fast decoherence is the fact that the polarization of the half-cycle pulse is nearly perpendicular to the quantization axis set by the static external electric field.  Each kick must change $m$ by $\pm 1$ allowing for many more states to be populated and small changes in the time between HCP's will result in dephasing between these states.

\MedianFigure

Fig. \ref{fig:contour}(d) shows the results of applying the above 1-D model to the initial state distribution shown in Fig. \ref{fig:contour}(b) measured when no half-cycle pulses have been applied. The predicted ionizing field distribution shown in Fig. \ref{fig:contour}(c) is calculated by assigning an electric field distribution to each principal quantum number centered at the expected field for red-shifted Stark states with a width proportional to the distance between adjacent state field ionization center values.  

To compare the model and experimental data in more detail, Fig. \ref{fig:median} shows the electric fields at which 10\%, 50\%, and 90\% of the atoms have ionized for a series of different initial states.  The agreement between the experimental data and the one free parameter model is quite good especially for atoms in higher $n$ states.   The lack of agreement for lower $n$ states, or equivalently populations that ionize at higher electric fields, in Fig. \ref{fig:median} is caused by a series of factors.  First at lower principal quantum numbers the 1-D approximation, even for a red-shifted Stark state, begins to break down.  Additionally half-cycle pulses applied to atoms in these states are more likely to drive atoms to states for which the 1-D approximation no longer applies.  In addition the width in electric field over which a state ionizes is proportional to the field at which it ionizes.  Thus for lower $n$ states the width is much greater resulting in a lower signal at a given field and increasing the sensitivity to noise and the exact details of how the conversion between state and ionizing field distributions is performed.  

\QHCPFigure

The maximum momentum transferred per pulse utilized in the model was calculated by a best fit to the data for the field at which 10\% of the total population was ionized including several initial state distributions as shown in Fig. \ref{fig:qhcp}(b).  Averaging over these points, we find a calculated momentum transfer per pulse of $q = 7.9\times 10^{-5}$.  From Eq. \ref{eq:t} and this calculated momentum transfer, there is less than a 1\% probability of an atom changing principal quantum number by $\pm 1$ from the action of a single pulse for states near $n \sim 50$ and a negligible probability to change by higher amounts.  Consequently this model behaves as a diffusive process and the spreading of population density is expected.  From the measured electric field of the half-cycle pulse we expect a maximum momentum transferred by a single pulse of $5\times 10^{-4}$ in atomic units.  This is a factor of 5 times more than the maximum calculated from the 1-D model fit as shown in Fig. \ref{fig:qhcp}(b).   However, correcting the calculated value for the reduced transition amplitudes when the HCP kicks the atom perpendicular to its quantization axis, we find $q \sim 3 \times 10^{-4}$ which is in close agreement.  

\section{Conclusion}

In summary, a train of half-cycle pulses produces large diffusive population transfers independent of the initial \textit{n}-state of the Rydberg atoms and operates regardless of the time structure creating Rydberg atoms.  These trains are thus ideal for the control and efficient de-excitation of Rydberg antihydrogen atoms as they meet all of the conditions required for compatibility with antihydrogen production discussed previously.  The ability to accurately model the population transfer demonstrated in this text utilizing analytical calculations of the inelastic form factor will allow for the efficient development of enhanced protocols utilizing, for example, chirped pulse trains to both enhance the efficiency of the transfer as well as to drive atoms toward a desired final state.  These protocols will depend on coherent population transfers and thus the rapid dephasing observed in our experiments must be reduced prior to their implementation.

Regardless of further developments, the population transfers demonstrated in this paper will greatly increase the probability of trapping antihydrogen atoms.  The 10\% increase in \textit{l}-mixed population with $n\leq 40$ has a radiative lifetime a factor of 4 shorter than that of $n=52$; furthermore, during the 150\microsecond\, that half-cycle pulses were applied, an additional 2\% of the population reached the ground state. For antihydrogen applications, this reduction from over a 4 ms to a 1 ms lifetime should increase the population that reaches the ground state before annihilating on a trap wall by over an order of magnitude based on the current ATRAP experimental configuration\cite{Gabrielse2008}.  

\begin{acknowledgments}
The authors acknowledge support by the Rowland Institute at Harvard. The emitters and detectors were produced at the Center for Nanoscale Systems (CNS), a member of the National Nanotechnology Infrastructure Network (NNIN), which is supported by the National Science Foundation.
\end{acknowledgments}

\end{document}